\documentclass[a4paper]{article}

\usepackage[pdftex]{graphicx}
\usepackage{amsmath}
\usepackage{authblk}
\usepackage{lineno}
\usepackage{url}

\title{
Determination of the light exposure on the
photodiodes of a new instrumented baffle for the
Virgo input mode cleaner end-mirror  
}

\author[1]{A. Romero}
\author[2]{A. Allocca}
\author[3]{A. Chiummo}
\author[1,4]{M. Mart\'{\i}nez}
\author[1]{Ll.M. Mir}
\author[5]{H. Yamamoto}
\date{\today}

\affil[1]{\emph{\small Institut de F\'{\i}sica d'Altes Energies (IFAE), Barcelona Institute of Science and Technology, Barcelona, Spain}}
\affil[2]{\emph{\small Universit\`a di Napoli `Federico II', Dipartimento di Fisica `Ettore Pancini', Complesso Universitario Monte S. Angelo, Napoli, Italy}}
\affil[3]{\emph{\small European Gravitational Observatory (EGO), Cascina, Pisa, Italy}}
\affil[4]{\emph{\small Institució Catalana de Recerca i Estudis Avançats (ICREA), Barcelona, Spain}}
\affil[5]{\emph{\small  LIGO laboratory, California Institute of Technology (Caltech), Pasadena, CA, US}}

\numberwithin{equation}{section}

\begin{document}

\maketitle

\begin{abstract}
As part of the upgrade program of the Advanced Virgo interferometer,  
the installation of new instrumented baffles surrounding the main test masses is foreseen. 
As a demonstrator, and to validate the technology, 
the existing baffle in the area of the input mode cleaner
end-mirror will be first replaced by a baffle equipped with photodiodes. 
This paper presents detailed simulations of the light distribution on the input mode cleaner baffle.
They served to validate the proposed layout of the sensors in the baffle, 
and determine the light exposure of the photodiodes under different scenarios
of the interferometer operations, 
in order to define mitigation strategies for preserving the detector integrity. 
\end{abstract}

\section{Introduction}
\label{sec:introduction}
Advanced Virgo is a power-recycled Michelson interferometer 
with 3 km long Fabry-Perot cavities in the two orthogonal arms~\cite{Acernese_2014}. 
Its next upgrade, 
named Advanced Virgo Plus (AdV+),
will occur in two phases.
The first phase, or Phase I, 
is currently taking place between the O3 and O4 observation runs, 
while Phase II will take place between the O4 and O5 observation runs~\cite{PhysRevLett.123.231108}.
Among several other improvements,
AdV+/Phase II foresees the installation of baffles instrumented 
with photosensors surrounding the main test masses.
The information provided by these photosensors will
improve the understanding of the stray light (SL) distribution at low angles in the interferometer,
detect the appearance of excited higher order modes which show up as modified patterns in the SL detected by the baffles,
open the possibility of monitoring the contamination of the mirror surfaces that leads to low-angle scattering,
and facilitate a more efficient pre-alignment and fine-tuning of the parameters of the interferometer after shutdowns and during operations.

The current LIGO~\cite{cQGaLigo} and KAGRA~\cite{Akutsu:16} interferometers
have photodiodes embedded in the baffles around the test masses
to ease the pre-alignment of the beam along the long arms.
These sensors also provide information on the amount of low-angle scattered light, 
but there are too few to supply its full distribution.
On the other hand, the Virgo baffles around the tests masses
are only equipped with aluminum targets to monitor the light scattered off of them.

As part of the Phase I upgrade, 
the replacement of the input mode cleaner (IMC) end-mirror and payload is being planned. 
This motivated the replacement of the current baffle by a new one instrumented with photodiodes (PDs), acting as a demonstrator of the selected technology. 
It will also serve to gain experience on operating such a new device within Virgo.  

\begin{figure}[htb]
	\centering
	\includegraphics[width=1.0\textwidth]{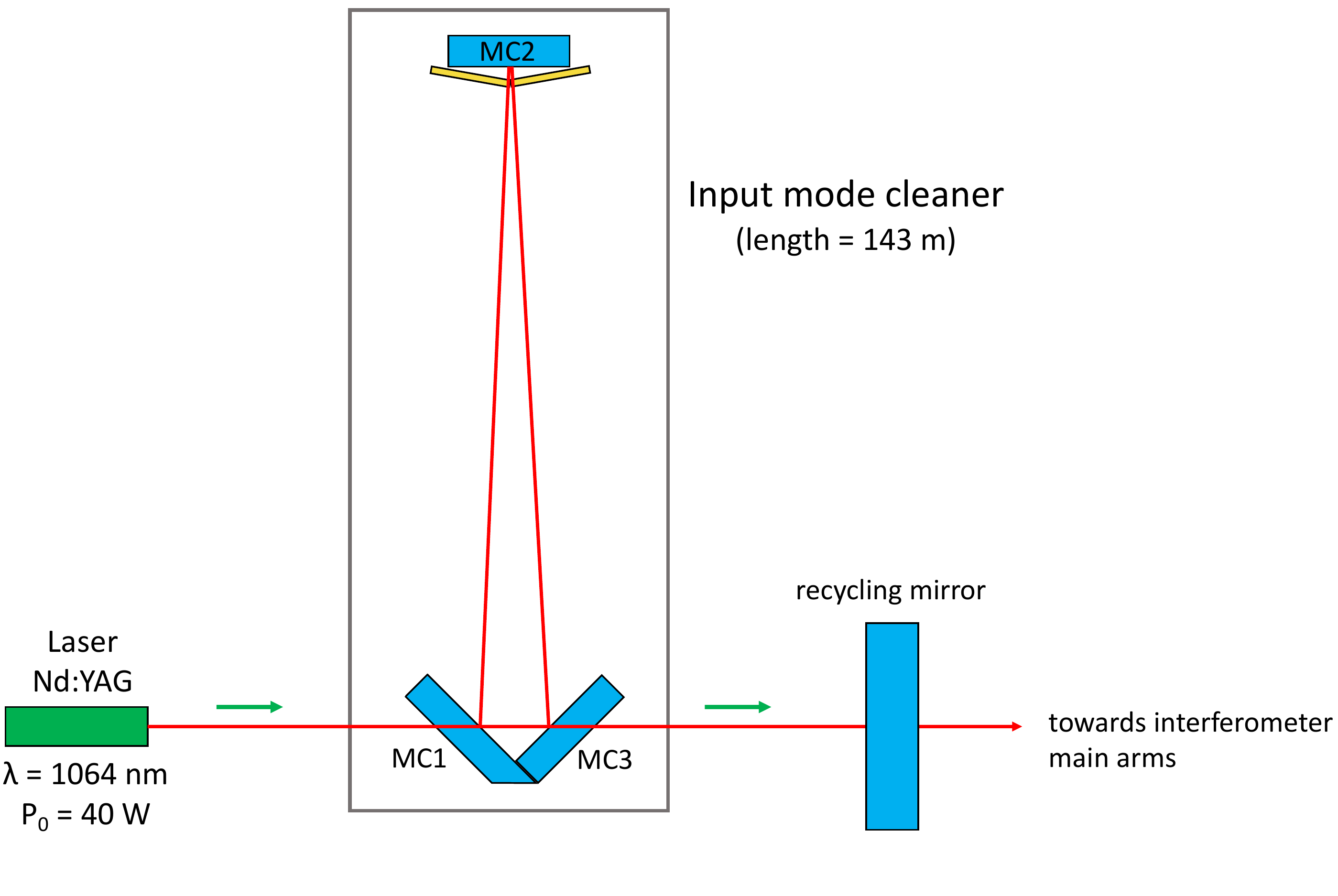}
	\caption{Sketch of the IMC cavity geometry.
    The instrumented baffle will be placed around the end-mirror (MC2).
    The yellow rectangles illustrate the position of the baffle.}
	\label{fig:IMC}
\end{figure}

\begin{figure}[htb]
	\centering
	\includegraphics[width=0.8\textwidth]{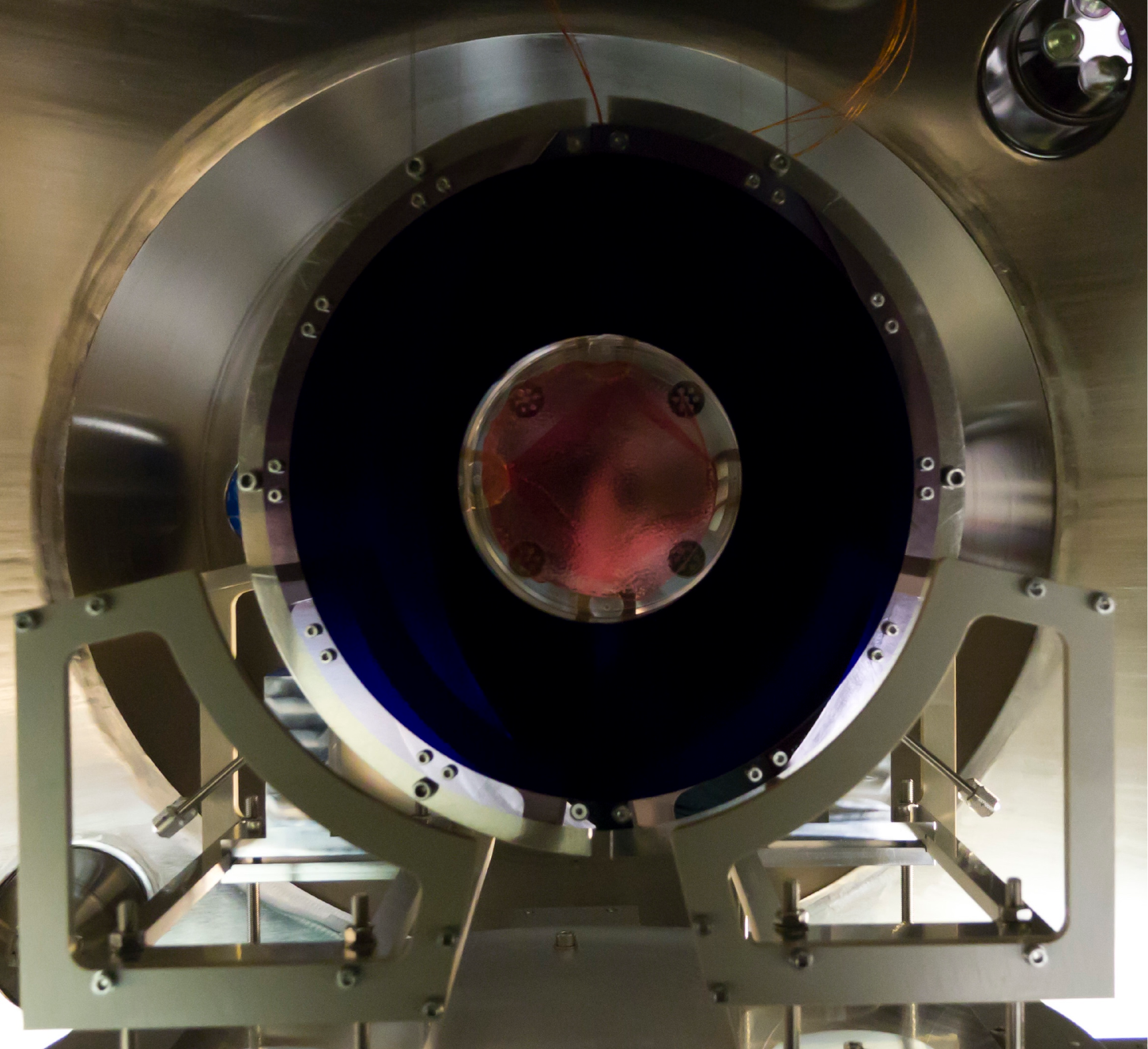} 
	\caption{Current non-instrumented baffle (dark ring) surrounding the IMC end-mirror (pink). Also visible is the safety frame at the bottom of the picture.}
	\label{fig:currBaffle}
\end{figure}

The IMC cavity is an in-vacuum triangular cavity with suspended optics, 
used for modal and frequency filtering of the laser beam before entering the interferometer.
Figure~\ref{fig:IMC} shows a schematic optical setup of the IMC,
where MC2 is the end-mirror
and MC1 and MC3 are the input and output mirrors, respectively.
The end-mirror has a radius of curvature of 187~m,
whereas the input- and output- mirrors are flat.
The half round trip length is approximately 143~m. 
Table~\ref{tab:IMC} summarises the IMC optical parameters relevant for
the calculations in this paper.
The complete information can be found in reference~\cite{Acernese_2014}.
Figure~\ref{fig:currBaffle} shows a picture of the current IMC end-mirror area, 
with the suspended mirror and the surrounding non-instrumented baffle~\cite{currentBaffle}.  

\begin{table}[htb]   
    \begin{center}
	\begin{tabular}{l|l|c} 
	\multicolumn{1}{c|}{Parameter} & \multicolumn{1}{c|}{Symbol} & \multicolumn{1}{|c}{Value}  \\ 
	\hline \hline
    Finesse             & $F$        & 1000 \\
    Free spectral range & $FSR$      & $1.04 \times 10^6$ Hz  \\
    MC1 transmissivity  & $T_{in}$   & $\simeq 2.5 \times 10^{-3}$   \\
    MC3 reflectivity    & $R_{out}$  & $\simeq 1$ \\
	\end{tabular}
	\caption{Parameters of the IMC needed for the calculations in this paper.\label{tab:IMC}}
	\end{center}
\end{table}

The new baffle is being instrumented with novel Si-based photosensors developed by Hamamatsu
and originally based on their S13955-01 series~\cite{hama},
with modifications in the packaging that render them  compatible with ultra-high
($10^{-9}$~mbars) vacuum conditions. 
The PDs are mounted on two large PCBs, 
which are not exposed to the light and include eight
temperature sensors distributed across their surface. 
The front-end electronics also includes protection measures against large laser doses, 
in order to preserve the integrity of the PDs.   

\begin{figure}[htb]
	\centering
	\includegraphics[width=0.6\textwidth]{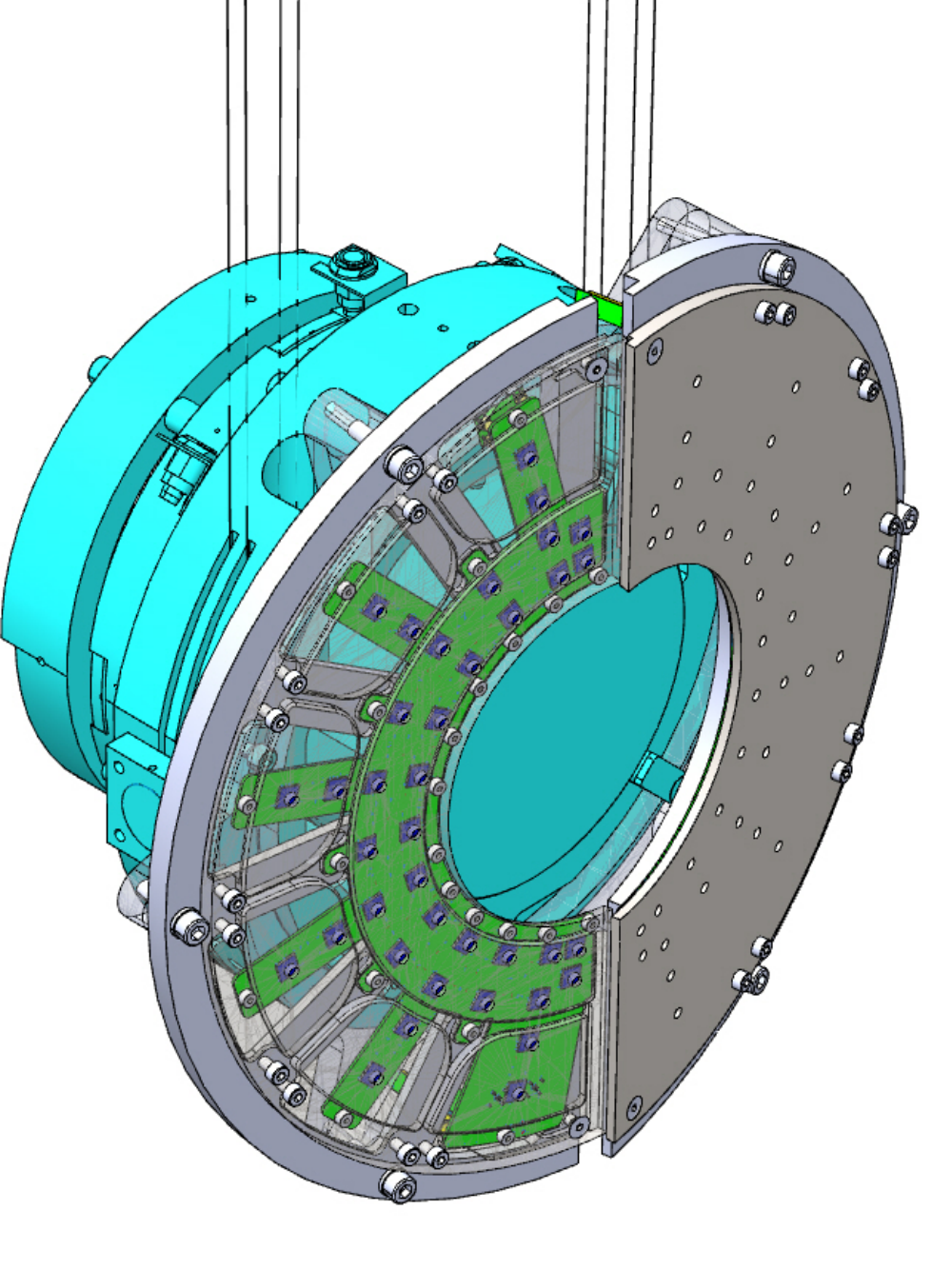}
	\caption{Design of the instrumented baffle surrounding the IMC end-mirror.}
	\label{fig:newBaffle1}
\end{figure}

The infrared light from the IMC cavity penetrates the holed mirror-polished 
stainless steel baffle reaching the PDs behind them. 
The distribution of the sensors on the baffle can be seen in figure~\ref{fig:newBaffle1},
and follows the pattern of light reaching the baffle as obtained
in the simulation studies shown in next section.
The largest amount of light is expected to be near the inner radius of the baffle,
and thus this region is populated densely with
52 PDs arranged in two concentric lines.
Twenty-four additional PDs cover the external region of the baffle for a total of 76 PDs.

The baffle, as a whole, 
is designed to preserve as much as possible the optical properties of the existing one 
in terms of reflectivity and total scattering 
to maintain the performance of the IMC cavity. 
In particular, 
both baffle and sensor surfaces are anti-reflecting coated and all
the components have been validated for ultra-high vacuum compatibility.
Furthermore,
the inner edge of the baffle and the edges of the holes
are produced with a small radius of curvature to prevent scattering.
Finally,
the weight and centre-of-gravity of the new baffle will be very similar to those of the current one,
to facilitate the installation in the existing suspension system.  
A publication with a detailed description of all the instrumental aspects 
of the new baffle is now in preparation. 
         
This paper presents simulation studies to determine the distribution
of light that reaches the IMC end-mirror and the surrounding baffle.
Different scenarios in the operation of the interferometer are considered,
which vary the amount of light illuminating the PDs. 
They include the nominal operation with the IMC cavity in resonance, 
the presence of small beam misalignments, 
and transient noise events leading to a severe displacement of the laser beam inside the cavity. 
Throughout this study an area of 0.49 cm$^2$ is used for a single PD, 
which corresponds to its effective sensitive area.
However,
since the PDs are partially obscured by conical holes of 0.4 cm of diameter, 
the power they will actually be exposed to 
will be in principle reduced by a factor of four.
The size and shape of the holes have been chosen such
that no light reaches the edges of the PDs.
All the results presented in the following sections are obtained for a laser 
input power of 40~W, as foreseen in O4. 
The highest power reached in O3 was 25 W.

\section{Input Mode Cleaner in Resonance}
\label{sec:simulations}
In this section we evaluate the amount of light that reaches the baffle 
when the IMC is in resonance, 
both under the assumption of perfect alignment and for misalignments 
compatible with the resonance state.
The fields in the IMC are calculated using a software package called SIS, 
Stationary Interferometer Simulation~\cite{Yamamoto_2006, fabry},
which was developed within the LIGO collaboration and made available to the 
gravitational waves community~\cite{SISmanual}.
This software calculates the stationary state field in an optical system by calculating 
its propagation and interaction with the optics using the Fresnel approximation of the Huygens integral. 
Fast Fourier transforms are used to accelerate the conversion of fields between the spatial domain
to the frequency domain during the field calculation.

All relevant IMC data, 
size, location and reflectivity of the mirrors, and apertures of the baffles in front of the mirrors,
are included in the simulation.
Of particular importance is the inclusion of the surface maps of the mirrors, 
which were measured at LMA with a spatial resolution of 0.35 mm. 
Since the incident angle of the field to MC2 is 0.02 degrees,
a reflected field is scattered back by MC2 to MC3. 
The power of this backward propagating field is around 0.1\% of the field propagating in the normal direction. 
Although this backward field is not explicitly included,
it is used to calculate the power of the forward field correctly.

In the implementation presented in this paper 
the physical properties of the baffle are not considered,
only its dimensions and its geometrical location surrounding the end-mirror,
where the field is evaluated.
Although this might be considered a crude approximation,
the simulation still provides sensible results given the fact 
that the amount of light illuminating the baffle is very small.

As discussed below, 
the results of the simulation rely on the detailed description of the mirror surface maps, 
the mirror apertures and interference effects, 
leading to non-Gaussian tails dominating the power spectrum away from the center of the mirror. 
On the other hand,
the power spectrum at the core of the mirror could be described, 
in first approximation, 
by a naive resonator model, 
leading to a Gaussian distributed spectrum as a function of the distance from the center of the mirror.


\subsection{Perfectly aligned cavity}
\begin{figure}[htb]
	\centering
	\includegraphics[width=0.48\textwidth]{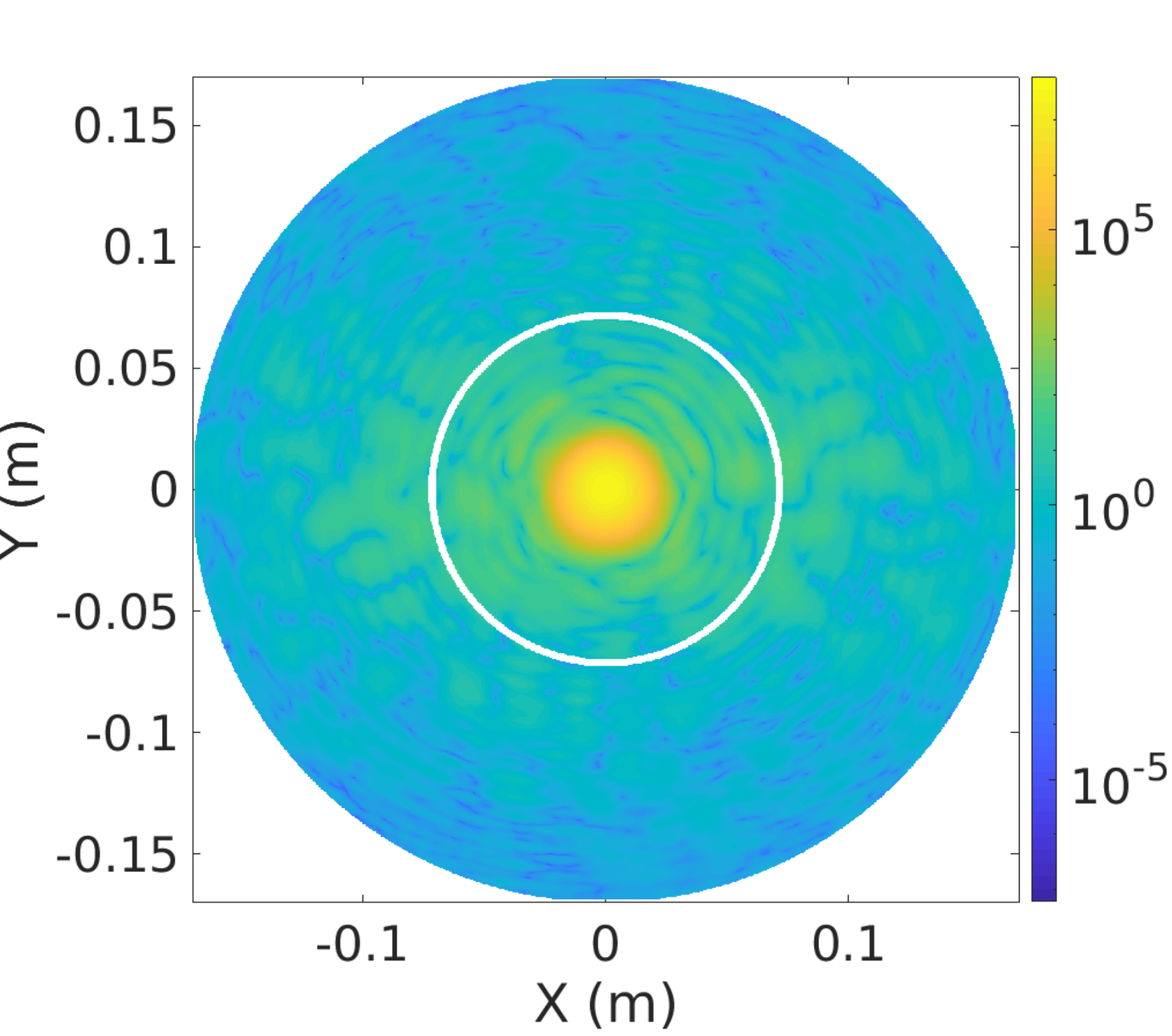}
    \includegraphics[width=0.48\textwidth]{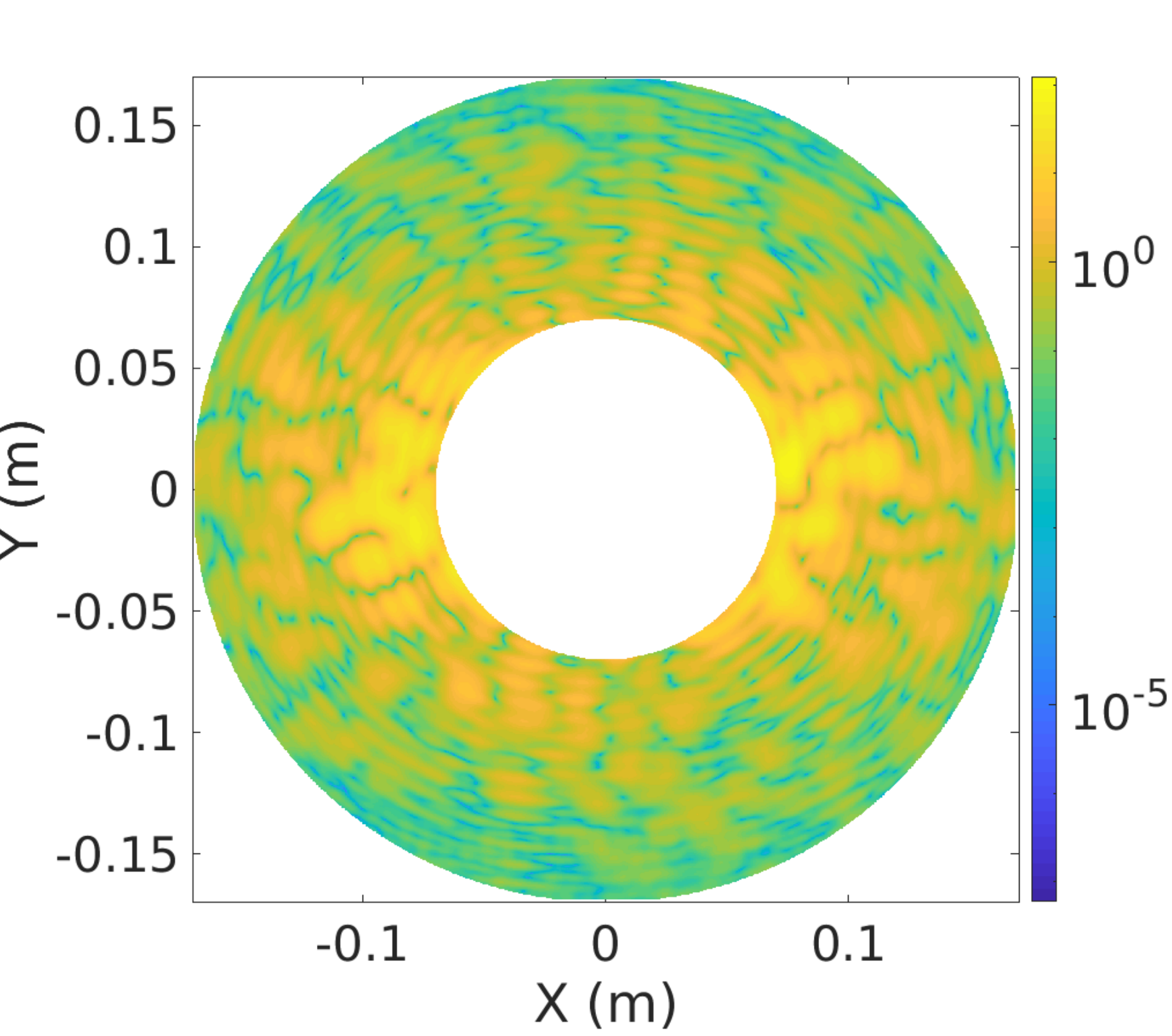}	
	\caption{2-D map of the power distribution 
	(in $\rm W/m^2$) 
	in the ensemble mirror plus baffle (left)
	and in the baffle surface only (right)
	for the cavity completely aligned. 
	The white line in the left plot indicates the inner radius of the baffle.}
	\label{fig:aligned}
\end{figure}

We first consider the IMC locked and perfectly aligned.
In this steady state condition the laser beam hits the center of the end-mirror. 
The power light distribution in the ensemble mirror plus baffle can be seen in figure~\ref{fig:aligned} left.
The total power reaching the mirror and the baffle, altogether, 
is of the order of $1.35\times 10^{4}$~W. 
Figure~\ref{fig:aligned} right shows the 
power reaching the baffle surface only, which amounts to 0.21~W,  
a $1.6\times 10^{-3}$~\% of the total power. 
The region in the baffle with the maximum light exposure is 
located in the horizontal axis near the right inner edge of the baffle,
and amounts to about $2\times 10^{-2}$~W over an area of 3.1~cm$^2$. 
Expressed in terms of sensors, 
this implies that a PD located in that region would receive a maximum dose of about 
$3.2\times 10^{-3}$~W, 
whereas a PD located in the outer part of the baffle,
at a radius of 17~cm from the center of the mirror,
would receive a power of the order of $3.2\times 10^{-5}$~W. 
The very low dark current for the PDs in the baffle,
at a maximum level of about 5000 pA, 
will allow detecting light power at the level of $10^{-5}$~W with more 
than three orders of magnitude of margin in the signal over noise ratio. 

\subsection{Misaligned cavity}
\begin{figure}[htb]
	\centering
	\includegraphics[width=0.55\textwidth]{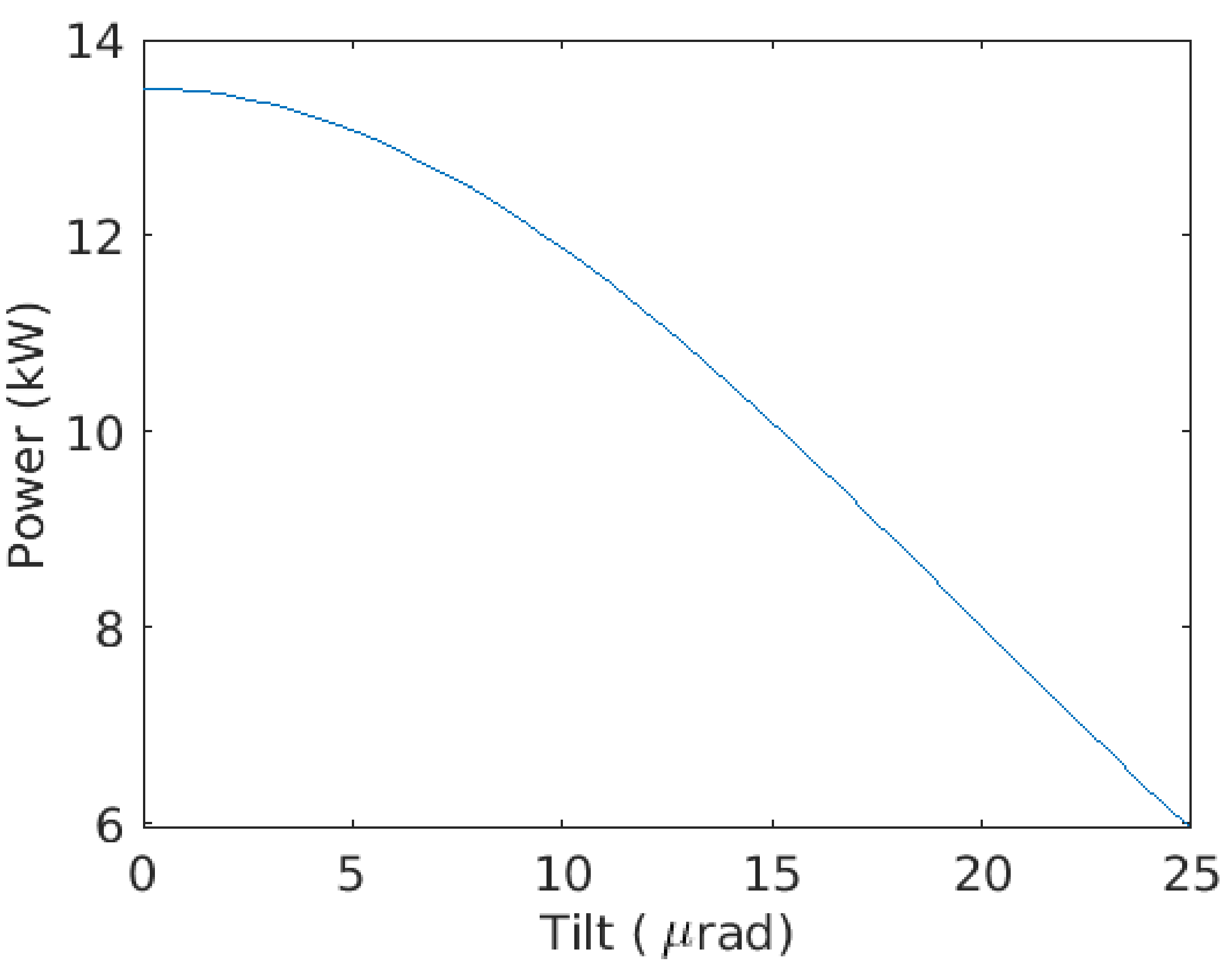}
	\caption{Total power (in W) reaching the mirror plus baffle ensemble as a function of the tilt applied to the end-mirror.}
	\label{fig:powerVSmisalignment}
\end{figure}

We now consider a scenario in which the IMC is misaligned,
but the misalignment is such that the cavity remains in resonance. 
The misalignment is implemented in the simulation via a tilt  
of the end-mirror with respect to its nominal position,
which results into a vertical displacement of the beam position.
Figure~\ref{fig:powerVSmisalignment} shows the power in the mirror plus baffle
ensemble as a function of the tilt angle $\alpha$ 
in the range 0 to 25~$\mu$rad. 
Predictably, the power decreases with increasing tilt.
As mentioned above, 
the baffles in the simulation are included as extensions of the mirrors.

\begin{figure}[htb]
    \centering
    \includegraphics[width=0.48\textwidth]{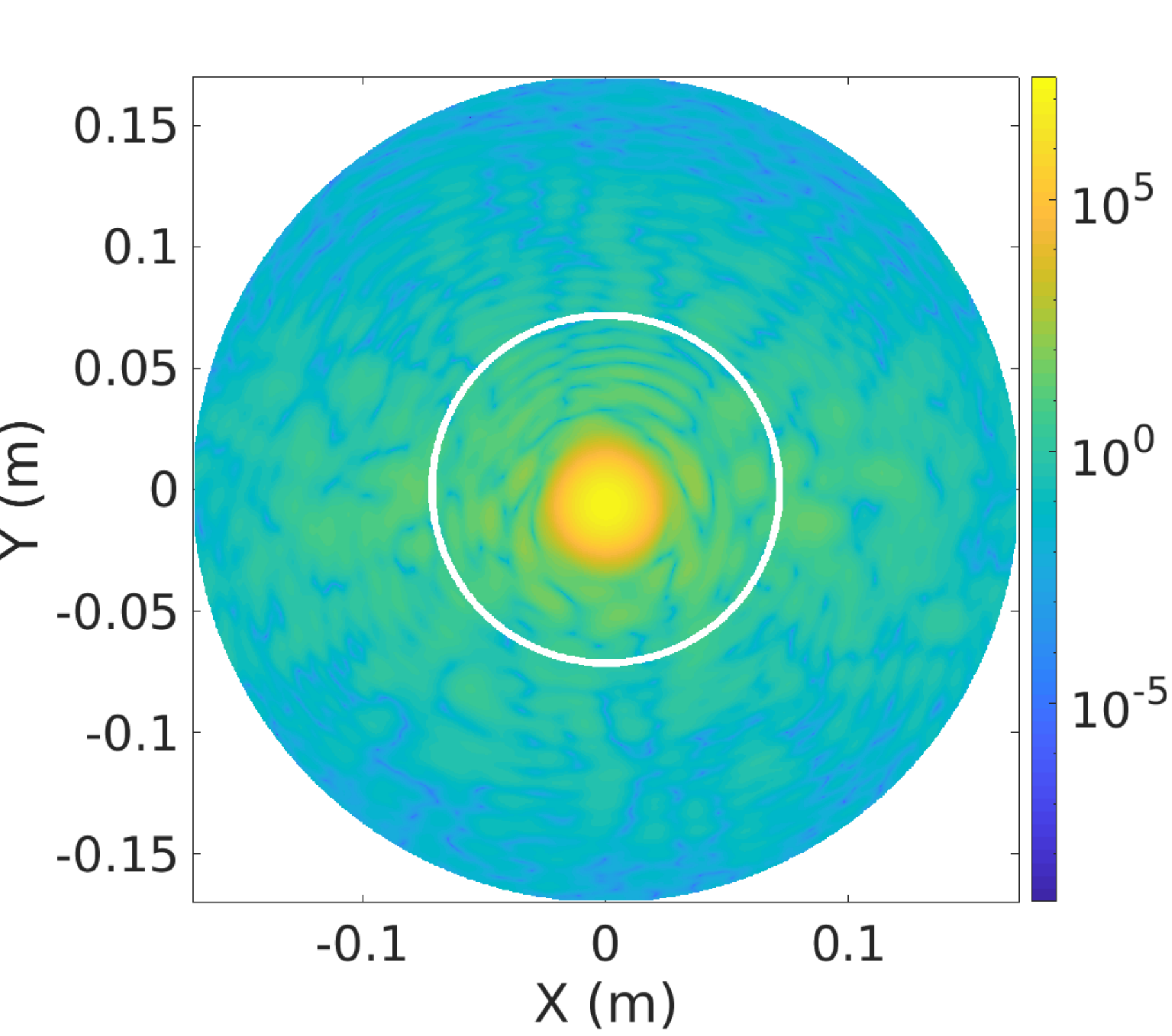}
    \includegraphics[width=0.48\textwidth]{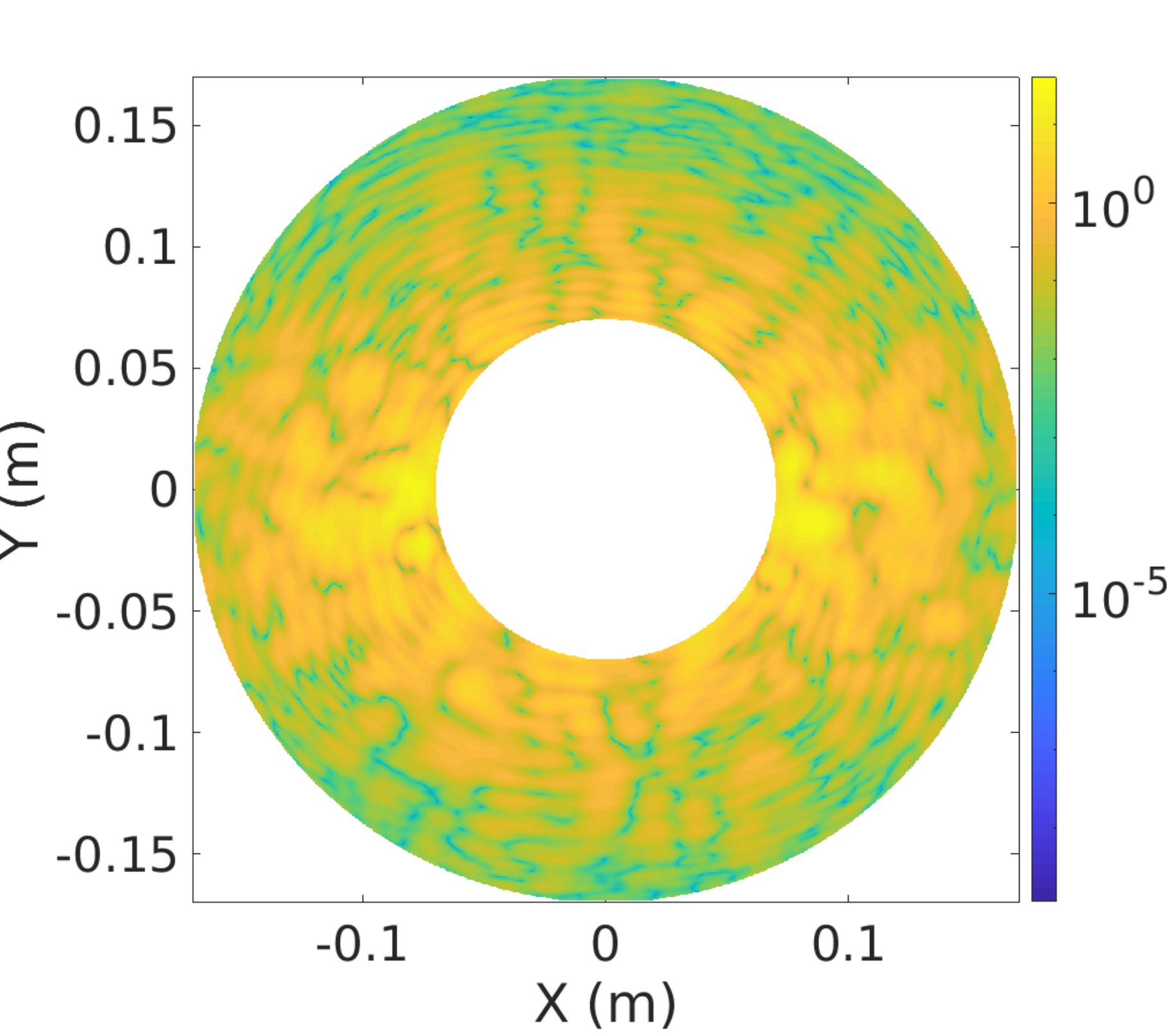}
    \caption{2-D map of the power distribution 
	(in $\rm W/m^2$) 
	in the ensemble mirror plus baffle (left)
	and in the baffle surface only (right)
    for the cavity on resonance but with a misalignment  
	$\alpha = 10~\mu$rad.
	The white line in the left plot indicates the inner radius of the baffle.}
    \label{fig:misaligned}
\end{figure}

As an example,
figure~\ref{fig:misaligned} left shows the power distribution for the case 
$\alpha$~=~10~$\mu$rad. 
The total power in the mirror and baffle ensemble becomes 
$1.19\times 10^{4}$~W,  
where only $0.17$~W (a $1.4\times 10^{-3}$~\% of the total power) illuminate the baffle itself. 
As discussed below, 
this is not a homogeneous effect. 
Although the consequence of the beam misalignment is an overall reduction of the power 
in the mirror plus baffle ensemble,  
the power reaching a particular region of the baffle can become larger. 

Figure~\ref{fig:misaligned} right shows the power distribution in the baffle surface only.
The region in the baffle with the maximum exposure receives a total of $8.9\times 10^{-3}$~W, 
distributed in an area of 1.4~cm$^2$,
in the same location as in the aligned case.
This translates into a maximum exposure of a single PD of about $3.0\times 10^{-3}$~W.
These results are dominated by uncertainties due to the limited accuracy in the description of the mirror maps, difficult to measure precisely over all spatial wavelengths.

\begin{figure}[htb]
    \centering
    \includegraphics[width=0.8\textwidth]{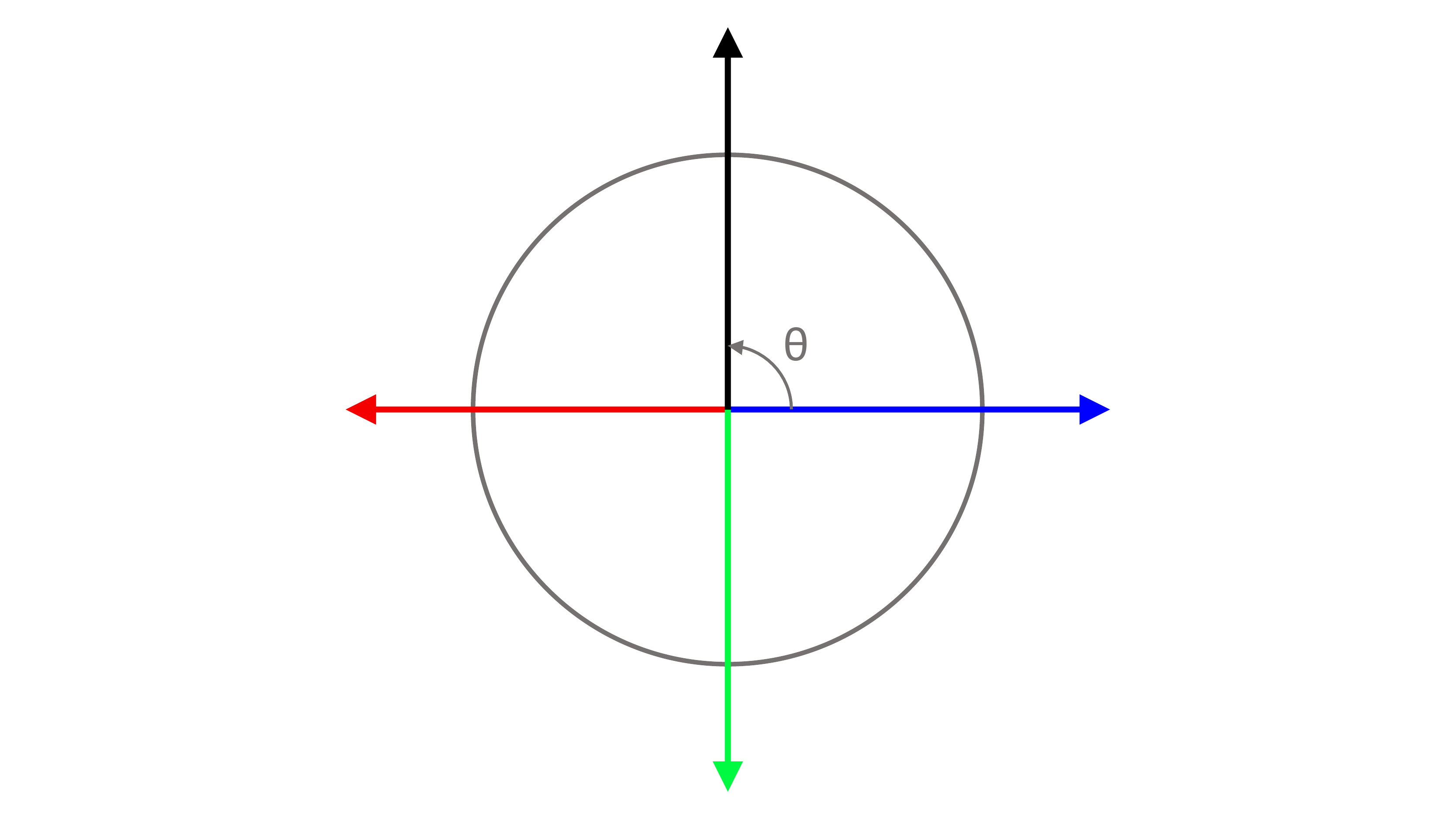}
    \caption{
Definition of $\theta$ and color code used in figure~\ref{fig:Radial}.}
    \label{fig:theta}
\end{figure}

\begin{figure}[htb]
    \centering
    \includegraphics[width=1.0\textwidth]{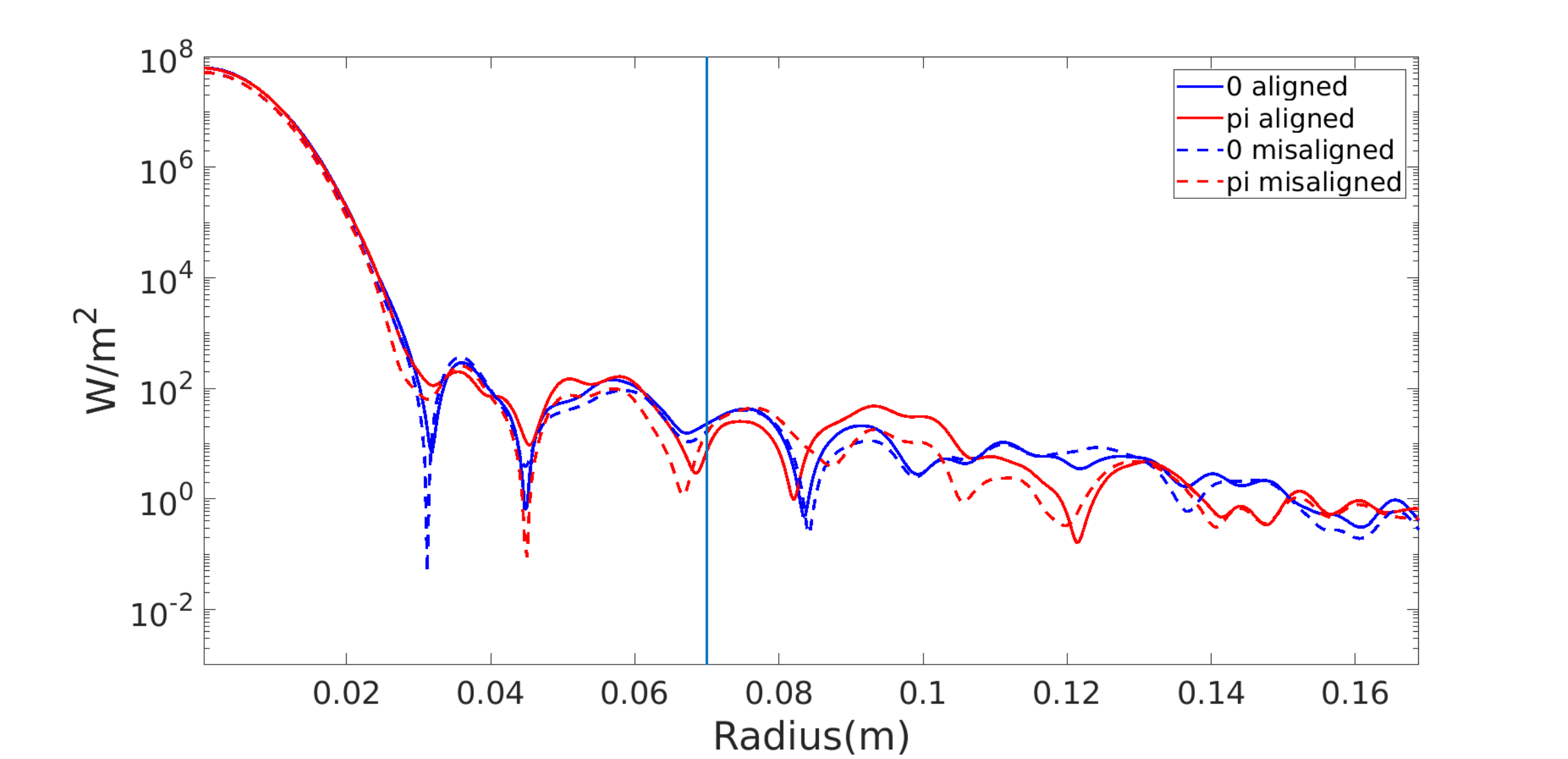}
    \includegraphics[width=1.0\textwidth]{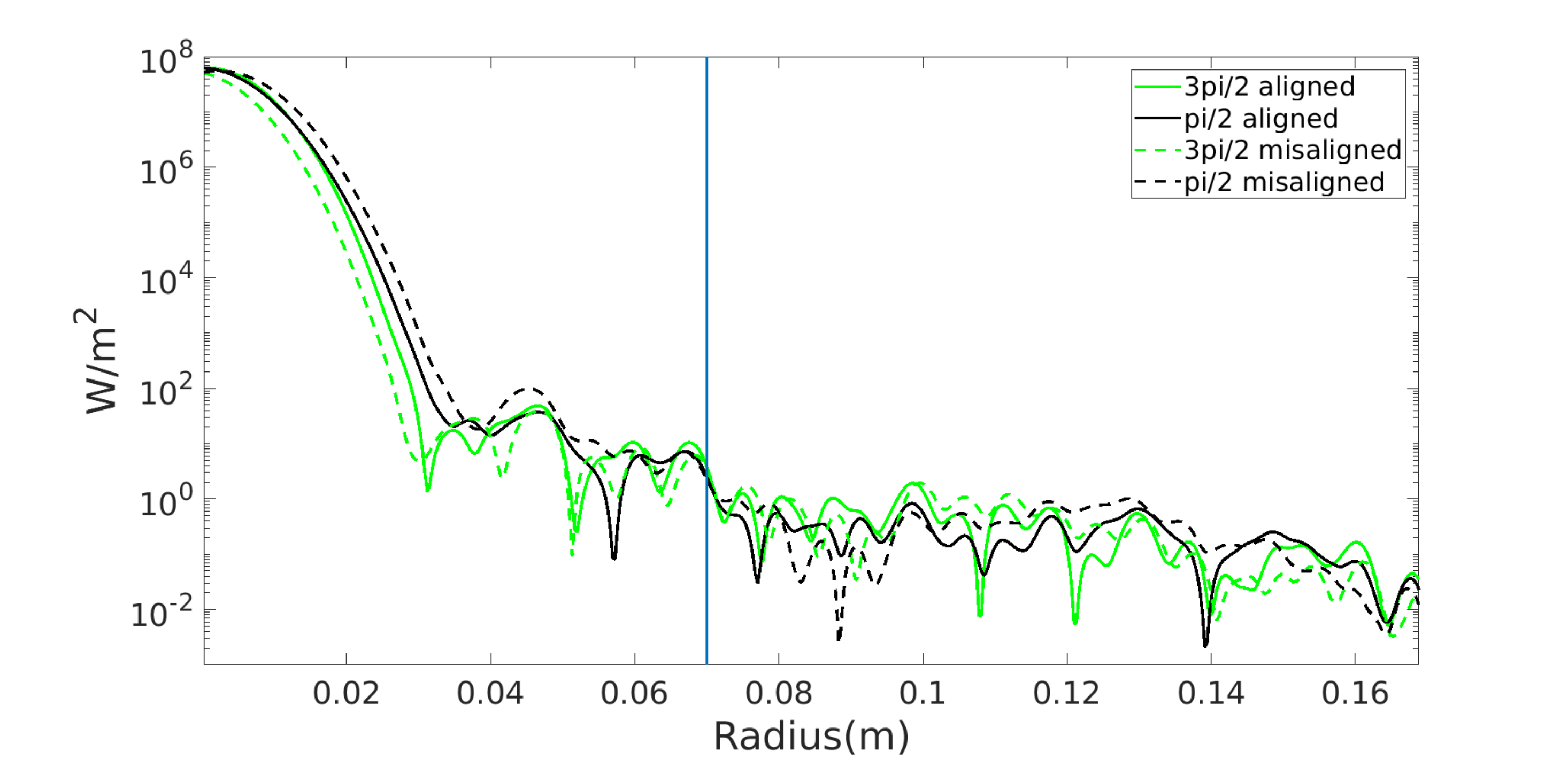}
    \caption{
Distribution of power (in $\rm W/m^2$) as a function of the distance from the center of the mirror at $\theta=0$ and $\theta=\pi$ (top)
and $\theta=\pi/2$ and $\theta=3\pi/2$ (bottom)
for the IMC cavity completely aligned and
with a misalignment of $\alpha = 10~\mu$rad.
The vertical line indicates the inner radius of the baffle.
The PDs are located at a distance of about 8 to 14 cm from the center of the mirror.}
    \label{fig:Radial}
\end{figure}

Given that the mirror has not perfect circular symmetry,
the radial distribution of the power depends on the angle $\theta$,
depicted in figure~\ref{fig:theta}.
For illustration purposes,
figure~\ref{fig:Radial} shows the power as a function of the radius 
for different values of $\theta$.
As anticipated, 
the power spectrum is approximately Gaussian-distributed at small radius 
and presents long non-Gaussian tails at large radius.
In the case of the aligned cavity scenario,  
slightly different spectra are determined for different $\theta$ values.  
When the cavity is misaligned, 
leading to a vertical displacement of the beam, 
no effect is observed in the bulk of the distribution at small radius for $\theta =0,\, \pi$, 
whereas a clear change is observed in the vertical direction for $\theta = \pi/2, \, 3\pi/2$. 
The changes due the misalignment in the non-Gaussian tails are more difficult to interpret since mirror surface maps, apertures and interference effects dominate, 
mixing vertical and horizontal effects.
Altogether, 
this illustrates the importance of instrumenting the area surrounding the mirrors to determine the accuracy of the simulation with data.  
For example, a simple resonator model implementing ideal mirrors with no defects would predict, 
in the presence of misalignments, 
a large relative increase in the power seen by the baffles.       

\clearpage

\section{Input Mode Cleaner out of Resonance}
\label{sec:calculations}
In this section we evaluate the amount of light that could illuminate the baffle 
in two extreme situations, first when the laser beam is so misaligned that the 
cavity loses its lock, 
and then in the case of severe displacements of the beam within the cavity 
caused by transient noise. 
In both cases we resort to analytical calculations to derive the power reaching the PDs.

\subsection{Misaligned cavity out of resonance}

For a extremely misaligned beam ($\alpha > 35~\mu$rad) the cavity loses its lock.  
In order to determine the level of exposure of a PD in such configuration,  
an analytical calculation is performed based on the
transmissivities and reflectivities of the mirrors of the cavity~\cite{Reitze:2019nwo}. 
For a beam of transverse area $A_{beam}$, 
the intensity in MC2, $I_{MC2}$, is evaluated as 

\begin{equation}
I_{MC2} = P_{in} \times T_{in} \times R_{out} / A_{beam},
\label{Iin}
\end{equation}

\noindent where $T_{in}$ is the transmissivity of MC1 and $R_{out}$ is the  reflectivity of MC3.
Using the values for $T_{in}$ and $R_{out}$ shown in table~\ref{tab:IMC}, 
and a beam diameter of 1~cm at the MC2,
$I_{MC2} = 1.3\times 10^{3}~\rm W/m^2$. 
Assuming a Gaussian beam illuminating directly a PD,
the maximum exposure it would receive becomes $2.1\times 10^{-2}$ W.
Laser-induced damage-threshold tests performed at the laboratory indicate 
that the PDs have a light power tolerance at least two orders of magnitude larger 
than the light dose expected to reach the sensors in each lock loss.   
 
\subsection{Transient noise, mechanical drift}

Finally, 
we have considered the scenario of a sudden cavity misalignment due to transient noise,
which would lead to a mechanical drift that could potentially result 
in exposing the PDs for a short period of time to the energy stored in the cavity.
The total energy stored in the cavity in nominal conditions can be expressed as 
$E_{\rm cav} = P_{\rm in} \times g \times \tau$~\cite{Reitze:2019nwo}, 
where $P_{\rm in}$ is the input power, 
$g$ is the gain of the cavity,  
defined in terms of its finesse $F$ as $g = {2 F}/{\pi}$, 
and $\tau$ is the decay time of the cavity
(average time that a photon stays in the cavity)
defined in terms of its finesse and free spectral range $FSR$
as $\tau = F / (2\pi \times FSR)$.
Using the values in table~\ref{tab:IMC},
$g\sim 640$, $\tau = 153~\mu$s, and $E_{\rm cav} = 3.9$~J.  

To translate the total energy in the cavity to the total power illuminating the baffle,  
the time response of the payload and suspension systems needs to be considered.  
It is expected that the feedback systems will need 10 ms to apply measures to correct the misalignment, 
bringing the power illuminating the baffle back to tiny values.  
During the 10 ms reaction time, however, 
the baffle is potentially exposed to a power not exceeding 390~W. 

To determine the maximum radiation that could hit a PD in this extreme case,
we consider a beam of Gaussian transverse profile with 390~W of amplitude
and 1~cm of size pointing to the center of a PD.
The power it would be exposed to is obtained
by integrating the Gaussian shape of the power distribution over the area of the PD, 
and results in a total power of about 130~W.
This is a rough estimation 
since a much more detailed analysis would be required to fully address this scenario.


\section{Summary}
\label{sec:summary}
Table~\ref{tab:sum} summarizes the values of the power reaching the IMC end-mirror and baffle ensemble,
the power reaching the baffle, 
and the maximum power that could impinge on a PD,
for the different scenarios discussed in the paper.
As explained in the introduction, 
even if the PDs are partially hidden by the baffle,
their total sensitive area has been used in the calculations.
Thus, the quoted values are conservative.

\begin{table}[htb]      
    \begin{center}
	\begin{tabular}{l|c|c|c} 
	\multicolumn{1}{c|}{Scenarios} 
	& Mirror plus baffle & Baffle & PD \\ 
	\hline \hline
    Resonance & $1.35 \cdot 10^{4}$ & 0.20 & $3.2\cdot 10^{-3}$ \\ 
    \hline
    Misaligned ($10~\mu\rm rad$)&  $1.19 \cdot 10^{4}$ & 0.17 & $3.0\cdot 10^{-3}$\\ 
    \hline
	Extremely misaligned & -- & -- & $2.1\cdot 10^{-2}$\\ 
	\hline
	Mechanical drift & 390 & -- & 130 (for 10 ms)\\ 
	\end{tabular}
	\caption{Values (in W) for the total
	power reaching the mirror and baffle ensemble,
	the power reaching the baffle and the maximum power that could impinge on a PD, 
	for the different scenarios discussed in the paper and an input laser power of 40~W.\label{tab:sum}}
	\end{center}
\end{table}

Due to the simplicity of the model used to simulate the baffle, 
the values shown in table~\ref{tab:sum} should be taken as a semi-quantitative estimate 
of the power in the baffle area in the different scenarios. 
Therefore, 
the data obtained with the instrumented baffle surrounding of the IMC end-mirror
will be used not only to determine the relevance of the PDs geometry
in the baffles to be placed around the test masses and the level of anticipated light doses, 
relevant for risk mitigation measures in the front-end electronics, 
but also to calibrate and further improve the simulation tools.

\section*{Acknowledgements}

The authors gratefully acknowledge the European Gravitational Observatory (EGO) and the Virgo Collaboration for providing access to the facilities. 
The LIGO Observatories were constructed by the California Institute of Technology and Massachusetts Institute of Technology with funding from the National Science Foundation under cooperative agreement PHY-9210038. The LIGO Laboratory operates under cooperative agreement PHY-1764464.
This work was  partially  supported   by  the Spanish MINECO   under   the grants
SEV-2016-0588 and PGC2018-101858-B-I00,  
some of which include ERDF  funds  from  the  European  Union.  
IFAE  is  partially funded by the CERCA program of the Generalitat de Catalunya.



\begin{thebibliography}{9}

\bibitem{Acernese_2014}
F. Acernese et al. (Virgo Collaboration),
\textit{Advanced Virgo: a second-generation interferometric gravitational wave detector},
Classical and Quantum Gravity 32 (2015) 024001.

\bibitem{PhysRevLett.123.231108}
F. Acernese et al. (Virgo Collaboration),
\textit{Increasing the Astrophysical Reach of the Advanced Virgo Detector via the Application of Squeezed Vacuum States of Light},
Phys. Rev. Lett. 123 (2019) 23.

\bibitem{cQGaLigo}
J. Aasi et al. (LIGO Collaboration),
\textit{Advanced LIGO},
Classical and Quantum Gravity 32 (2015) 074001.

\bibitem{Akutsu:16}
T. Akutsu et al., 
\textit{Vacuum and cryogenic compatible black surface for large optical baffles in advanced gravitational-wave telescopes},
Opt. Mater. Express 5 (2016) 6.

\bibitem{currentBaffle} 
Nikhef@Virgo Mechanical Technology Department,\\
\textit{Input Mode Cleaner and Dihedron},\\
\texttt{https://www.nikhef.nl/pub/departments/mt/projects/virgo/\\
inputmodecleaner.html},
2012.

\bibitem{hama}
Hamamatsu,
\textit{Data sheet of Si photodiode S13955-01},\\
\texttt
{https://www.hamamatsu.com/resources/pdf/\\
ssd/s13955-01\_etc\_kspd1087e.pdf}, 2020.

\bibitem{Yamamoto_2006}
H. Yamamoto et al.,
\textit{Simulation tools for future interferometers},
Journal of Physics: Conference Series 32 (2006) 061.

\bibitem{fabry}
J. Wenxuan, A. Effler and V. Frolov,
\textit{Physical and statistical analysis of scatter in Fabry-Pérot Arm Cavity of Advanced LIGO"},
LIGO-T1800224-v1, 2018,
\texttt{https://dcc.ligo.org/LIGO-T1800224/public}.

\bibitem{SISmanual}
H. Yamamoto,
\textit{SIS (Stationary Interferometer Simulation) Manual},
LIGO-T070039-v8, 2013,
\texttt{https://dcc.ligo.org/LIGO-T070039-v8/public}.

\bibitem{Reitze:2019nwo}
P. Saulson,
\textit{Advanced Interferometric Gravitational-wave Detectors},
Edited by D. Reitze and H. Grote, Volume I, WSP, 2019,
isbn: 978-981-314-607-5. 
  
\end{thebibliography}
\end{document}